\documentclass[11pt,dvips]{article}

\usepackage{epsfig,times}

\usepackage{picinpar}
\usepackage{wrapfig}
\usepackage{floatflt}

\makeatletter
\renewcommand{\section}{\@startsection{section}{1}{0in}
	{0.4\baselineskip}{0.1\baselineskip}{\Large\bf}}
\renewcommand{\subsection}{\@startsection{subsection}{2}{0in}
	{0.25\baselineskip}{-\baselineskip}{\large\bf}}
\renewcommand{\subsubsection}{\@startsection{subsubsection}{3}{0in}
	{0.1\baselineskip}{-\baselineskip}{\normalsize\bf}}
\makeatother

\begin{document}

%
\makeatletter\newcommand{\ps@icrc}{
\renewcommand{\@oddhead}{\slshape{OG.1.3.04}\hfil}}
\makeatother\thispagestyle{icrc}

%
%

\begin{center}
{\LARGE \bf The Anisotropy of Cosmic Ray Arrival Direction around 10$^{18}$eV}
\end{center}

\centerline{ (AGASA Collaboration)}

\begin{center}

{\bf 
N. Hayashida$^{1}$,
K. Honda$^{2}$,
N. Inoue$^{3}$,
K. Kadota$^{4}$,
F. Kakimoto$^{4}$,
S. Kakizawa$^{5}$,
K. Kamata$^{6}$,\\
S. Kawaguchi$^{7}$,
Y. Kawasaki$^{8}$,
N. Kawasumi$^{9}$,
E. Kusano$^{10}$,
Y. Matsubara$^{11}$,
K. Mase$^{1}$,
T. Minagawa$^{1}$,\\
K. Murakami$^{12}$,
M. Nagano$^{13}$,
D. Nishikawa$^{1}$,
H. Ohoka$^{1}$,
S. Osone$^{1}$,
N. Sakaki$^{1}$,
M. Sasaki$^{1}$,\\
M. Sasano$^{1}$,
K. Shinozaki$^{3}$,
M. Takeda$^{1}$,
M. Teshima$^{1}$,
R. Torii$^{1}$,
I. Tsushima$^{9}$,\\
Y. Uchihori$^{14}$,
T. Yamamoto$^{1}$,
S. Yoshida$^{1}$,
and H. Yoshii$^{15}$} \\

\bigskip

{\it
$^1$Institute for Cosmic Ray Research, University of Tokyo, Tokyo 188-8502, Japan\\
$^2$Faculty of Engineering, Yamanashi University, Kofu 400-8511, Japan \\
$^3$Department of Physics, Saitama University, Urawa 338-8570, Japan \\
$^4$Department of Physics, Tokyo Institute of Technology, Tokyo 152-8551, Japan\\
$^5$Department of Physics, Shinshu University, Matsumoto, Japan\\
$^5$Nishina Memorial Foundation, Komagome, Tokyo 113-0021, Japan\\
$^6$Faculty of General Education, Hirosaki University, Hirosaki 036-8560, Japan\\
$^7$Department of Physics, Osaka City University, Osaka 558-8585, Japan\\
$^8$Faculty of Education, Yamanashi University, Kofu 400-8510, Japan\\
$^9$KEK, High Energy Accelerator Organization, Tsukuba 305-0801, Japan\\
$^{10}$Solar-Terrestrial Environment Laboratory, Nagoya University, 
Nagoya 464-8601, Japan\\
$^{11}$Nagoya University of Foreign Studies, Nissin, Aichi 470-0131, Japan\\
$^{12}$Department of Applied Physics and Chemistry, Fukui University of Technology, Fukui 910-8505, Japan \\
$^{13}$National Institute of Radiological Sciences, Chiba 263-8555, Japan \\
$^{14}$Department of Physics, Ehime University, Matsuyama 790-8577, Japan \\
}

\end{center}

\begin{center}
{\large \bf Abstract\\}
\end{center}
\vspace{-0.5ex}

Anisotropy in the arrival directions of cosmic rays
around 10$^{18}$eV is studied
using data from the Akeno 20 km$^2$ array and 
the Akeno Giant Air Shower Array (AGASA), 
using a total of about 216,000 showers observed over 15 years
above $10^{17}eV$.
In the first harmonic analysis, we have found significant anisotropy
of $\sim$ 4 \% around 10$^{18}$eV, corresponding to a chance probability
of $\sim 10^{-5}$ after taking the number of independent 
trials into account.
With two dimensional analysis in right ascension and declination,
this anisotropy is interpreted as an excess of showers near
the directions of the Galactic Center and the Cygnus region.
This is a clear evidence for the existence of the galactic cosmic ray
up to the energy of $10^{18}eV$. Primary particle which contribute this anisotropy
may be proton or neutron.

\vspace{1ex}

\section{Introduction:}

Searches for anisotropy
in the arrival directions of high energy cosmic rays 
have been made by many experiments so far and 
the arrival direction distribution
of cosmic rays is found to be quite isotropic
over a broad energy range.  In the previous report (Hayashida et al. 1998),
we have reported the evidence of the anisotropy correlated with our
galaxy at $10^{18}$eV.
This result means the existence of galactic cosmic rays up to the
energy of $10^{18}eV$ and can be interpreted by two models, 
one is the proton diffision model and the other, the superposition 
of neutron emission from the galactic sources.
It is worthwhile to note that the neutron's life time become 
comparable with the scale of the galaxy at this energy range.
Here, we obtained more significant results by adding
data upto April 1999.

The Akeno Giant Air Shower Array (AGASA) consists of 111 
scintillation detectors of 2.2 m$^{2}$ area each,
which are arranged with 1km spacing over 100 km$^2$ area.
Akeno is located at latitude 35$^{\circ}$ 47'$N$
and longitude 138$^{\circ}$ 30'$E$ at an
average altitude of 900 m above sea level.
Details of the AGASA array are described in Ohoka et al. 1997.
Data from the 20 km$^2$ array (Teshima 1986)
are included in this analysis.  
The typical angular resolution is 3 degrees for 10$^{18}$eV cosmic
rays.

\section{Results:}

\begin{figwindow}[1,r,%
{\mbox{\epsfig{file=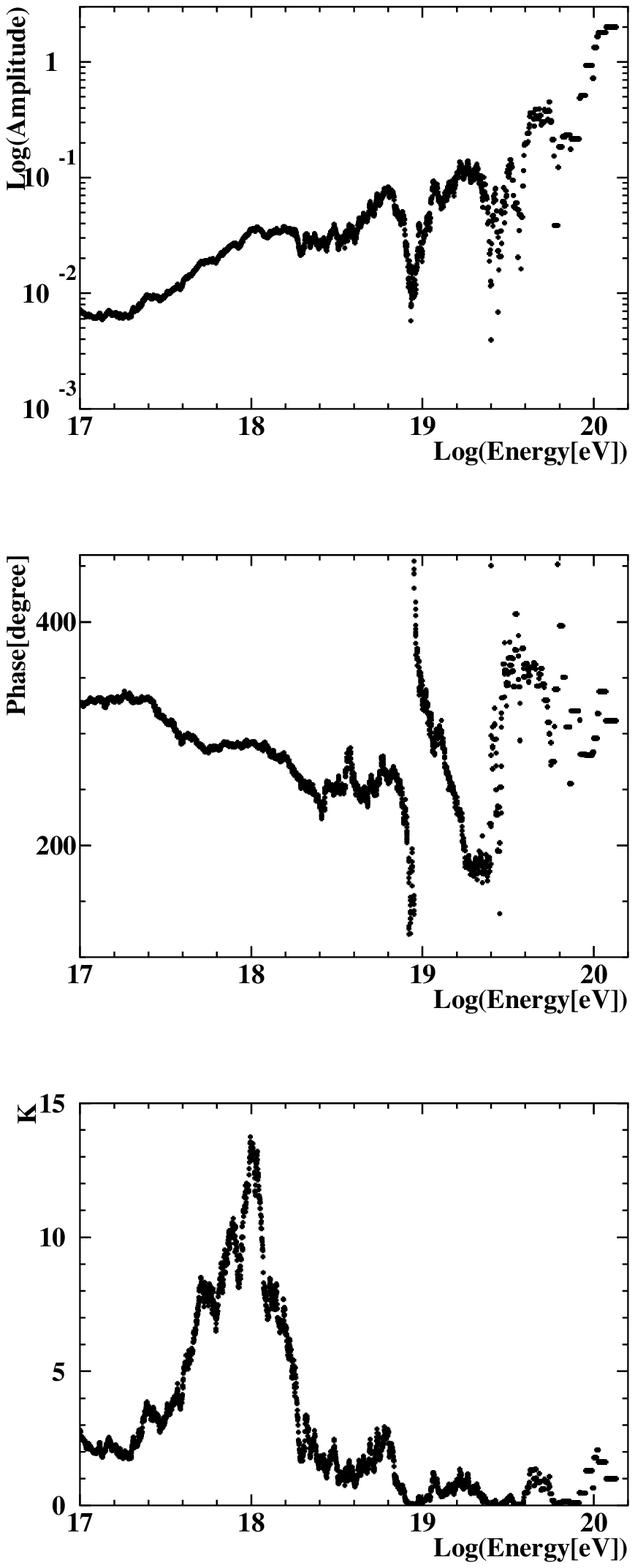,width=3.0in}}},%
{The first harmonic analysis. 
The amplitude, the phase, peak direction in right ascension, 
and the significance k are shown as a function 
of primary energy threshold, respectively.
Each point is obtained by summing over 
events with more than the corresponding energy. }]
The showers are selected with the following conditions.
The core is inside the array, the number of hit detectors is $\geq 6$,
and the reduced $\chi^{2}$ in determining the arrival direction
and the core position is less than 5.0.
All events with zenith angles $\leq 60^{\circ}$ are used 
in the present analysis.
About 216,000 events remain after these Selection.
Results of first harmonic analysis in right ascension
are shown in Figure 1.
The amplitude, the phase (peak direction in right ascension), 
and the significance $k$ are shown as a function of primary energy threshold.
Each point is obtained by summing over events with more than the corresponding 
energy. The parameter $k$ follows the probability distribution of exp(-k), if
we assume the random sample.
$k$ $\sim$ 14 around 10$^{18}$eV can be seen in the bottom plot.
It is surprisingly high, corresponding
to a chance probability of 10$^{-6}$. 
We have searched for the energy bin width which gives the maximum $k$
-value, and find that the region 10$^{18.0}$eV - 10$^{18.4}$eV 
gives the maximum $k$-value of 15.1. 
This means the showers which contribute to 
the anisotropy are distributed in the energy range of 0.4 decade.

We also listed the results in the differential bins with energy
ranges of a factor of two from 1/2EeV to 8EeV in Table 1,
for the comparison with the world data.
According to this table, the chance probability is estimated to be
$\sim 10^{-5}$ by taking the number 
of independent trials into account.

In searching for anisotropy,
rates from different regions on the celestial sphere are compared.
Therefore uniform observation time in right ascension
is quite important in this analysis. We have carried out several tests
on the systematic effect(see Hayashida et al. 1998), however, we found
the observation is quit uniform and the systematics are negligeble
small for the present $\sim 4 $\% amplitude.

In Figures 2, the arrival direction distributions in equatorial
coordinates are shown.
They show the ratio of the number of observed event to the expected one,
and the statistical significance of the deviations from the
expectation.
Here, the energy region of 10$^{18.0}$ $\sim$ 10$^{18.4}$eV is selected 
which maximize the harmonic analysis $k$-value.
We can not observe events with declination less than -25$^{\circ}$,
as long as we use showers with zenith angles less than 60$^{\circ}$.
In this figure, we have chosen a circle of 20$^{\circ}$ radius to evaluate
the excess. The expected event density inside this circle 
are evaluated using the average of event rate at the same declination band.

\end{figwindow}

\begin{table}
\caption{The first harmonic analysis in right ascension.}
\centerline{
\begin{tabular}{lcrrrll}\hline 
Bin & Energy Range/EeV &  \#   & Amplitude[\%] & Phase & k    & P$_{prob}$ \\
\hline
E4  &  1/2 - 1.0       & 56658 & 0.5     & 272    & 0.42 & 0.65	\\
E5  &  1.0 - 2.0       & 29207 & 4.2     & 297   & 12.9 & 2.5E-6\\
E6  &  2.0 - 4.0       & 10129 & 2.0     & 256   & 1.10 & 0.33	\\
E7  &  4.0 - 8.0       &  2769 & 3.3     & 256   & 0.76 & 0.46	\\
\hline
\end{tabular}
}
\end{table}

In the significance map with beam size of $20^\circ$, 
a $4.5 \sigma$ excess (obs./exp. = 506/413.6) near 
the Galactic Center region can be seen. 
In contrast, near the direction of anti-Galactic Center
we can see a deficit in the cosmic ray intensity ($ -4.0 \sigma$).
An event excess from the direction of the Cygnus region 
is also seen in the significance map with 3.9 $\sigma$
(obs./exp. = 3401/3148).

\section{Discussion}

An anisotropy of amplitude 4\% around 10$^{18}$eV was found
in first harmonic analysis. 
With a two dimensional map, we can identify this
as being due to event excesses of $4.5 \sigma$ and $3.9 \sigma$ 
near the galactic center and the Cygnus region, respectively.
The observed anisotropy seems to be correlated with the galactic structure.

Such anisotropy has not been observed by
previous experiments.  Since the latitudes of the Haverah Park and Yakutsk
are around 60 degrees, the direction of significant excess in
the present experiment near the galactic center can
not be observed by those experiments and hence a significant
amplitude in harmonic analysis might be absent in their data.
Fly's Eye data show the enhancement at similar energy range(Bird et al.1999).

One possible explanation of the anisotropy reported here is due to
the propagation of cosmic ray protons.
The observed regions of excess are directed toward the galactic plane and
seem to be correlated with the nearby spiral arms.
However, the direction of anisotropy need not point toward the
nearby galactic arm, since scattering is diffusive in the leaky box model.
According to the Monte Carlo simulation by Lee and Clay 1995,
a proton anisotropy of 10\% $\sim$ 20\% amplitude
is expected at RA $\sim$ 300$^{\circ}$
using an axisymmetric concentric ring model
of the galactic magnetic field with interstellar turbulence
of a Kolmogorov spectrum.  The source 
 distribution is assumed to be uniform within the galactic disk
and both a non-random and turbulent magnetic halo with
various field strengths are taken into account.
If the observed anisotropy is due to protons, we can estimate 
the proton abundance as to be about 20\% $\sim$ 40\% of  
all cosmic rays, by comparing our result of 4\% amplitude
 with their simulation.

Another possible explanation is that the anisotropy is due to neutron 
primary particles.
Neutrons of 10$^{18}$eV have a gamma factor of 10$^9$ and
their decay length is about  10 kpc. Therefore they can propagate
linearly from the galactic center without decaying. 
The accelerated heavy nuclei should interact 
with the ambient photons or gases in the acceleration region, 
and spill out neutrons. 
The produced neutrons can escape freely from the acceleration site. 
In this scenario, the heavy dominant 
chemical composition below 10$^{18}$eV (Gaisser et al. 1993) and the lack of 
anisotropy below 10$^{18}$eV (due to the short
neutron lifetime) can be naturally explained.

More accumulation of the data, observation in the southern hemisphere, 
and the determination of energy spectrum in the excess region are important to
confirm the experimental result and to discriminate
two possibilities.

\section*{Acknowledgment}
We are grateful to Akeno-mura, Nirasaki-shi, Sudama-cho, Nagasaka-cho,
Takane-cho and Ohoizumi-mura for their kind cooperation.
The authors also wish to acknowledge the valuable help by other members
of the Akeno Group in the maintenance of the array.

\newpage

\begin{figure}[h]
\centerline{\epsfig{file=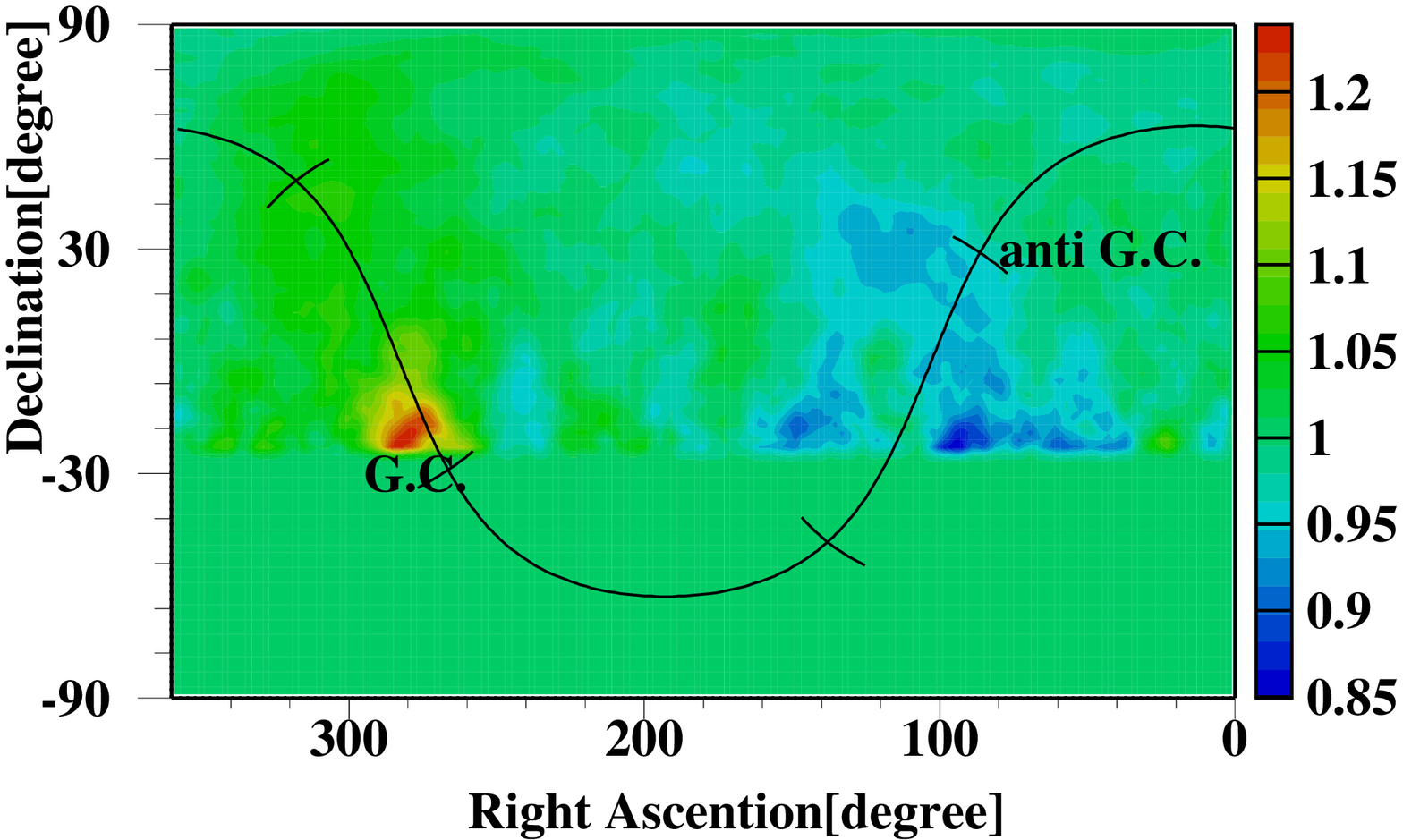,height=3.0in}}
\centerline{\epsfig{file=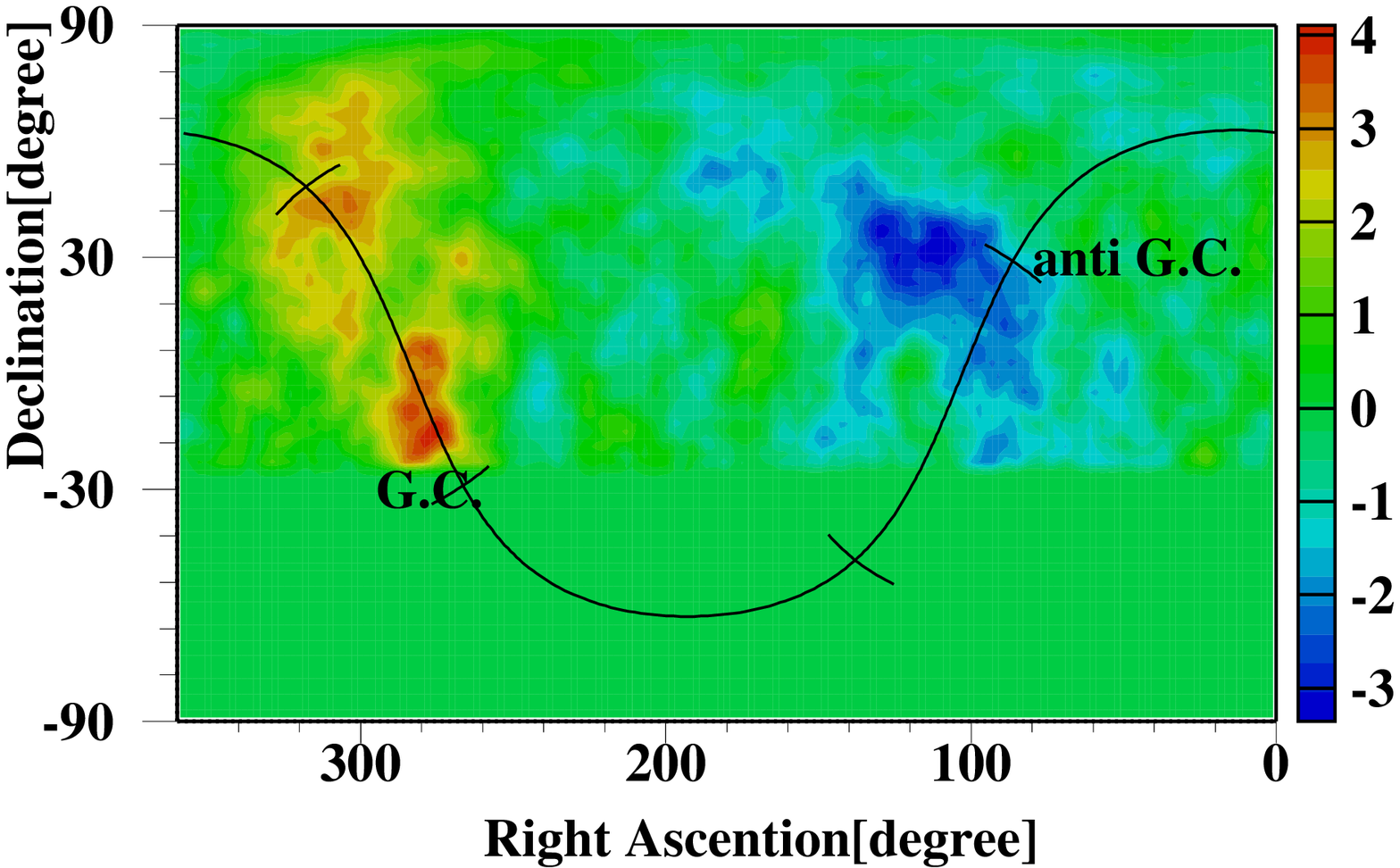,height=3.0in}}
\caption {The ratio of the number of observed events to the expected one
(top pannel)
and the statistical significance of the deviations (bottom pannel) 
are shown on the equatorial coordinate.}
\end{figure}

\vspace{1ex}
\begin{center}
{\Large\bf References}
\end{center}
%
Bird, D. et al. 1999, ApJ 551, 739 \\
Gaisser, T.K. et al. 1993, Phys. Rev. D47, 1919.\\
Hayashida, N. et al. 1998, astro-ph/9807045, Astrop. Phys. 10-4, 303.\\
Lee, A.A. and Clay, R.W. 1995, J. Phys. G: Nucl. Part. Phys. 21, 1743. \\
Ohoka, H. et al. 1997, Nucl.Istr. Meth. A385, 268. \\
Teshima, M. et al. 1986, Nucl.Instr.and Meth. A247, 399. \\

\end{document}